# New Schottky-gate Bipolar Mode Field Effect Transistor (SBMFET): Design and Analysis using Two-dimensional Simulation

M. Jagadesh Kumar[1] and Harsh Bahl

Department of Electrical Engineering,

Indian Institute of Technology, Delhi,

Hauz Khas, New Delhi – 110 016, INDIA.

*Abstract*- A new Schottky-gate Bipolar Mode Field Effect Transistor (SBMFET) is proposed and verified by two-dimensional simulation. Unlike in the case of conventional BMFET, which uses deep diffused $p^+$-regions as the gate, the proposed device uses the Schottky gate formed on the silicon planar surface for injecting minority carriers into the drift region. The SBMFET is demonstrated to have improved current gain, identical breakdown voltage and ON-voltage drop when compared to the conventional BMFET. Since the fabrication of the SBMFET is much simpler and obliterates the need for deep thermal diffusion of P+-gates, the SBMFET is expected to be of great practical importance in medium-power high-current switching applications.

*Index Terms*- Bipolar Mode Field Effect Transistor (BMFET), Schottky Gates, conductivity Modulation, two-dimensional (2-D) modeling.

# I. Introduction

MOS switches used in high-current medium-power switching applications, have good switching characteristics but suffer from high ON resistance. However, devices such as insulated gate bipolar transistor (IGBT) and bipolar mode field effect transistor (BMFET) use drift region conductivity modulation to derive very small ON resistance and hence negligible conduction losses [1-4]. BMFETs are also used very commonly in optical applications as an optically controlled switch [5-8] and have been reported both on GaAs [9-11] and silicon [12-22]. The most commonly studied BMFET devices have N-type epitaxial drift region and $P^+$-gates. When the $P^+$-gate is reversed biased, the BMFET operates in unipolar mode and the drain current is controlled similar to that in a JFET. The bipolar mode of operation starts when the $P^+$-gate is forward biased with respect to the source resulting in a significant hole injection into the N-drift region. The presence of this hole-electron plasma in the drift region leads to a substantial conductivity modulation resulting in a negligible saturation voltage (typically 0.2 V for Si) and, hence, a small ON resistance. However, the conventional BMFETs have deep $P^+$ gate junctions, which require long and expensive thermal cycles for fabrication.

The main purpose of this paper is to examine if these deep diffused $P^+$-gates can be replaced by Schottky gates while maintaining the bipolar mode operation. Schottky contacts, although very widely used in high-speed devices [23-30], are typically not opted for bipolar mode operation because of the general perception that Schottky contacts are usually dominated by majority carrier injection with no or negligible minority carrier injection. For the first time, we demonstrate that using a metal with an appropriate work function and choosing a very lightly doped N-drift region, the Schottky gate can be made to operate in a bipolar mode resulting in a new Schottky gate Bipolar Mode Field Effect Transistor (SBMFET). The Schottky gate in the proposed SBMFET obliterates the requirement of the deep thermal diffusion needed for the $P^+$ gate in a conventional BMFET and, therefore, results in a much simpler fabrication since the Schottky gate can be easily formed by metal sputtering on the planar silicon surface. In this work, using two-dimensional numerical simulation [31], we demonstrate that the Schottky-gate BMFET (SBMFET) exhibits a comparable or better performance when compared to the conventional silicon $P^+$-gate BMFETs in terms of breakdown voltage, ON-voltage drop and current gain while its fabrication method is much simpler.

## II. Bipolar Mode Operation of Schottky Gate

The cross-sectional views of the conventional BMFET and the SBMFET are shown in Fig. 1. The device parameters ($P^+$-gate junction depth, epilayer thickness and its doping, channel width, source, drain and $P^+$ gate peak doping) are chosen based on reported results in literature [12-15] and are tabulated in Table I. Two-dimensional numerical simulations were performed using MEDICI [31] to obtain the DC characteristics of BMFET and SBMFET. The various models activated in the simulations are Fermi-Dirac distribution for carrier statistics, Klaassen's unified mobility model for dopant-dependent low-field mobility, analytical field dependent mobility for high electric field, ionization rate model for impact ionization and Shockley-Read-Hall (SRH) and Klaassen Auger recombination models for minority carrier recombination lifetime. For simulating the Schottky gate junction properties, standard thermionic emission model is used.

The electron and hole concentrations in the N-drift region of the SBMFET are shown in Fig. 2 for different bias conditions. Fig. 2(a) shows the thermal equilibrium hole and electron concentration at zero bias condition i.e. at $I_G = 0$ µA/µm and $V_{DS} = 0$ V. It can be seen that the concentration of electrons at this bias is $2 \times 10^{13}$ cm$^{-3}$ which is equal to the epilayer doping and the hole concentration is approximately $7 \times 10^6$ cm$^{-3}$. Fig. 2(b) shows the concentration of holes and electrons in the channel for $I_G = 0.4$ µA/µm and $V_{DS} = 0.5$ V. On forward biasing the Schottky gate, i.e. for a positive gate current, the concentration of holes and electrons in the channel rises resulting in a conductivity modulation of the drift region. This clearly demonstrates the ability of the Schottky gate to inject excess holes into the N-drift region [32]. When the drain voltage is increased to $V_{DS} = 5$ V at $I_G = 0.4$ µA/µm, the holes are pushed deeper towards the source making the low resistance conductivity modulated region shorter as shown in Fig. 2(c). The presence of the conductivity modulated region and its variation with the drain voltage can also be observed by calculating the electric field in the drift region as shown in Fig. 3. It can be seen that for $V_{DS} = 0.5$ V, the low electric field region (i.e. the conductivity modulated region) is longer when compared to $V_{DS} = 5$ V.

The gate metal of the SBMFET plays a major role in facilitating the valance electron injection into the metal and should be chosen carefully. Otherwise, our simulation results have shown that plasma formation will not take place in the channel and the device will behave like a resistor without any saturation in the drain current with increasing drain voltage. The key to successful operation of the device is the formation of correct Schottky barrier. This is formed only when the work function of the metal ($\Phi_m$) is greater than that of the semiconductor ($\Phi_s$). Metals with work function lower than that of semiconductor form an ohmic contact and are thus unsuitable as gate metals. Thus Nickle ($\Phi_m = 5.15$ V), Platinum ($\Phi_m = 5.65$ V) are suitable, while Aluminum ($\Phi_m = 4.28$ V), Chromium ($\Phi_m = 4.5$ V) and Titanium ($\Phi_m = 4.33$ V) are unsuitable.

In order to understand the physics of the device, particularly the effect of Schottky interface, energy band diagrams of the Metal gate-N epilayer junction simulated using MEDICI are shown in Fig. 4 for the thermal equilibrium case and for a finite forward bias $V_{fb}$. i.e. apply a positive voltage to the metal gate with respect to the N-type substrate. If a gate metal of appropriate work function is used (i.e $\Phi_m > \Phi_s$), under forward bias conditions, the electrons from the valence band of the semiconductor also start flowing into the metal. This effect is equivalent to holes being injected into the semiconductor [32]. The injection of holes in the N-Epilayer region creates excess minority carrier holes. To maintain quasi-neutrality, the majority carrier electrons also get accumulated along with the minority carrier holes thus forming the conductivity-modulated region or hole-electron plasma region. It has been verified by simulations that the concentration of holes and electrons in the epilayer rises to nearly $2 \times 10^{15}$ /cm$^{-3}$. The increase in the concentration of electrons from the initial doping of $2 \times 10^{13}$/ cm$^{-3}$, indicates the formation of low resistance conductivity region. From the band diagram we also notice that under forward bias conditions $E_C$-$E_{Fnfb}$ is smaller than the $E_C$-$E_{Fn0}$ under no bias condition. The epilayer however remains electrically neutral. The concentration of electrons reduces to a low value at the metal-semiconductor interface as we move from the semiconductor to the metal side while that of the hole concentration increases.

Fig. 5 shows a comparison of the output characteristics of the BMFET with the SBMFET for different gate metals listed above. The device shall work for all metals with $\Phi_m > \Phi_s$ and does not depend on a particular metal or process employed. Our simulation results also indicate that for metals such as Aluminum, Chromium and Titanium, no plasma formation takes place in the channel and hence they are not suitable as the gate metals.

# III. Simulation of DC Characteristics and Discussion

*A. Output Characteristics*

The simulated output characteristics of the SBMFET are shown in Fig. 6 for different gate currents. It can be seen that as the drain voltage is increased, the slope of the drain current curve reduces. However, with increasing drain voltage, the drain current continues to increase, just as in the case of a conventional BMFET, since the plasma region, which is responsible for the conductivity modulation of the drift region, is pushed away from the drain terminal. Due to the conductivity modulation of the drift region populated by holes and electrons, the saturation voltage is, however, extremely small.

*B. DC Current Gain*

The current gain curves for BMFET and SBMFET for $V_{DS}$ = 5 V are shown in Fig. 7 in which we can see that the SBMFET exhibits a higher current gain compared to the conventional BMFET. It has been shown that for a given gate current, as the gate doping increases, the epilayer gets filled by a higher carrier density, which decreases the ON-resistance, and results in an increased drain current [15]. The doping concentration in the $P^+$ gate is always less than the free carrier concentration in the Schottky metal gate. Therefore, as the simulated electric field profile for the SBMFET and BMFET devices at $I_G$ = 0.4 µA/µm and $V_{DS}$ = 5 V shown in Fig. 8 demonstrates, the SBMFET has a better conductivity modulated region than the BMFET resulting in a higher drain current for a given gate current. As a result the SBMFET exhibits a higher current gain than the conventional BMFET.

A comparison of the current gain for the conventional BMFET and the SBMFET for different epilayer thicknesses is shown in Fig. 9 by varying the drift region thickness (W) from 20 µm to 56 µm. It is seen that the current gain reduces with increasing drift region thickness for both the devices since the drain current decreases with increasing drift region thickness at a fixed drain voltage. However, it is to be noticed that the SBMFET exhibits a slightly higher current gain than the conventional silicon BMFET for any given drift region thickness.

*C. Blocking Voltage*

The blocking voltage variation with epilayer thickness is shown in Fig. 10 for both the BMFET and the SBMFET. It can be seen that the blocking voltage increases linearly with increasing epilayer thickness. An increase in epilayer thickness allows the depletion region to expand and therefore can support a larger reverse bias at the drain terminal. The blocking voltage of the SBMFET is comparable to the conventional BMFET except for large values of epilayer thickness. However this is not a drawback as practical BMFETs are seldom used with a large epilayer thickness due to the low current gain at these thicknesses. It is thus clear that there is a trade-off between the current gain and the blocking voltage depending on the choice of the epilayer thickness.

It may be pointed out that the gate-to-drain breakdown voltage mainly determines the maximum blocking voltage of a BMFET device [35]. The schottky barrier between the gate and drain in a SBMFET breaks down slightly earlier than the P-N junction in a conventional BMFET. The SBMFET thus exhibits a slightly lower blocking voltage than a BMFET.

# IV. Conclusion

Bipolar mode field effect transistors are an ideal choice for high-current medium power switching applications. However, the conventional BMFETs need deep diffused $P^+$ gates to inject minority carriers into the drift region for conductivity modulation. In this paper, for the first time, we have proposed a new Schottky-gate BMFET (SBMFET) in which the Schottky gate, when forward biased, injects minority holes into the N-drift region. Our simulation results using

MEDICI demonstrate that the SBMFET exhibits an improved current gain while its breakdown voltage and ON-state voltage are similar to that of a conventional BMFET. Since the SBMFET obliterates the need for deep $P^+$ gates, but exhibits improved performance when compared to the conventional BMFET, it is expected to be of practical importance in many switching applications.

**Figure Captions**

Figure 1     Cross sectional view of (a) the conventional BMFET with deep diffused $P^+$-gates and (b) SBMFET with Schottky gates.

Figure 2     Hole and Electron concentration in the channel from source to drain at (a) $I_G = 0$ µA/µm and $V_{DS} = 0$ V, (b) $I_G = 0.4$ µA/µm and $V_{DS} = 0.5$ V and (c) $I_G = 0.4$ µA/µm and $V_{DS} = 5$ V.

Figure 3     Electric field variation in the channel for $V_{DS} = 0.5$ V and 5 V.

Figure 4     Energy band diagram of the gate- epilayer junction in a SBMFET at (a) zero gate-source bias and (b) when the gate is forward biased.

Figure 5     Comparison of output characteristics of BMFET and SBMFET using the following gate metals: Platinum ($\Phi_m=5.65$ V), Nickel ($\Phi_m=5.15$ V), Aluminum ($\Phi_m=4.28$ V), Chromium ($\Phi_m=4.5$ V) and Titanium ($\Phi_m=4.33$ V).

Figure 6     Output characteristics of SBMFET for different gate currents.

Figure 7     Current gain curves of BMFET and SBMFET for $V_{DS} = 5$ V.

Figure 8     Electric field variation in the channel for SBMFET and conventional BMFET at $V_{DS} = 5V$

Figure 9     Comparison of Current gain of BMFET and SBMFET with varying epilayer thickness for $I_G = 0.4$ µA/µm and $V_{DS} = 5$ V.

Figure 10     Blocking voltage variation of BMFET and SBMFET versus epilayer thickness. Gate is reverse biased at –10 V to ensure that the device is in OFF state.

Table I. Device parameters used for simulation of Conventional BMFET and SBMFET

| Device Parameter | BMFET | SBMFET |
|---|---|---|
| Epilayer thickness, W | 20-56 µm | 20-56 µm |
| Channel width, d | 31 µm | 31 µm |
| Source doping depth | 4 µm | 4 µm |
| $P^+$ Gate junction depth | 12 µm | --- |
| Gate Metal | --- | Nickel ($\Phi_m=5.15$ V) |
| Epilayer doping | $2 \times 10^{13}$ cm$^{-3}$ | $2 \times 10^{13}$ cm$^{-3}$ |
| Peak source doping | $1 \times 10^{20}$ cm$^{-3}$ | $1 \times 10^{20}$ cm$^{-3}$ |
| Peak gate doping | $1 \times 10^{17}$ cm$^{-3}$ | --- |
| Peak drain doping | $1 \times 10^{18}$ cm$^{-3}$ | $1 \times 10^{18}$ cm$^{-3}$ |

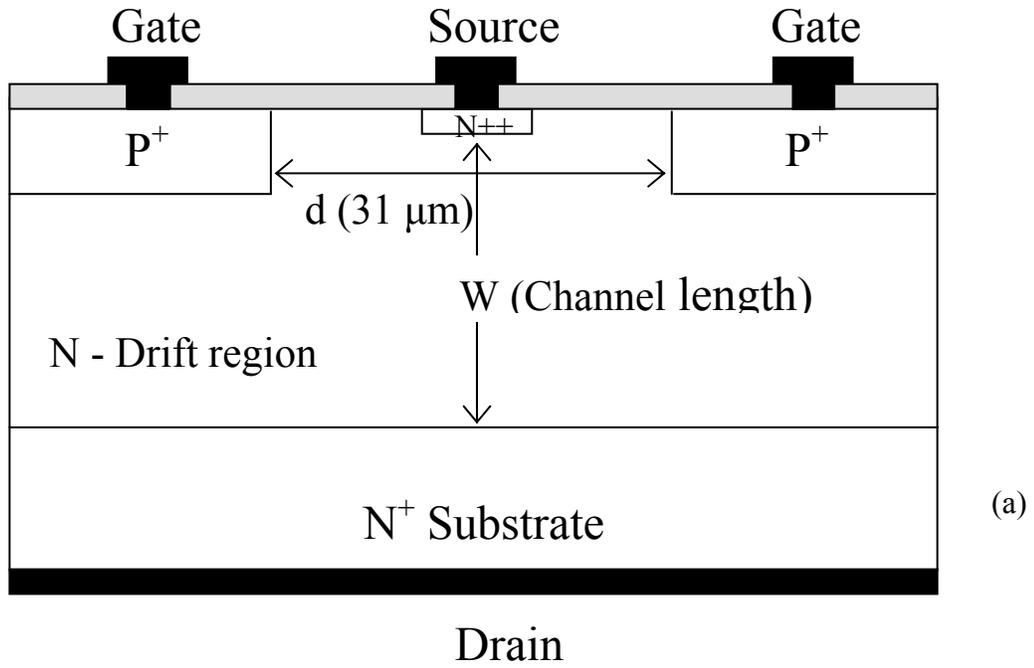

(a)

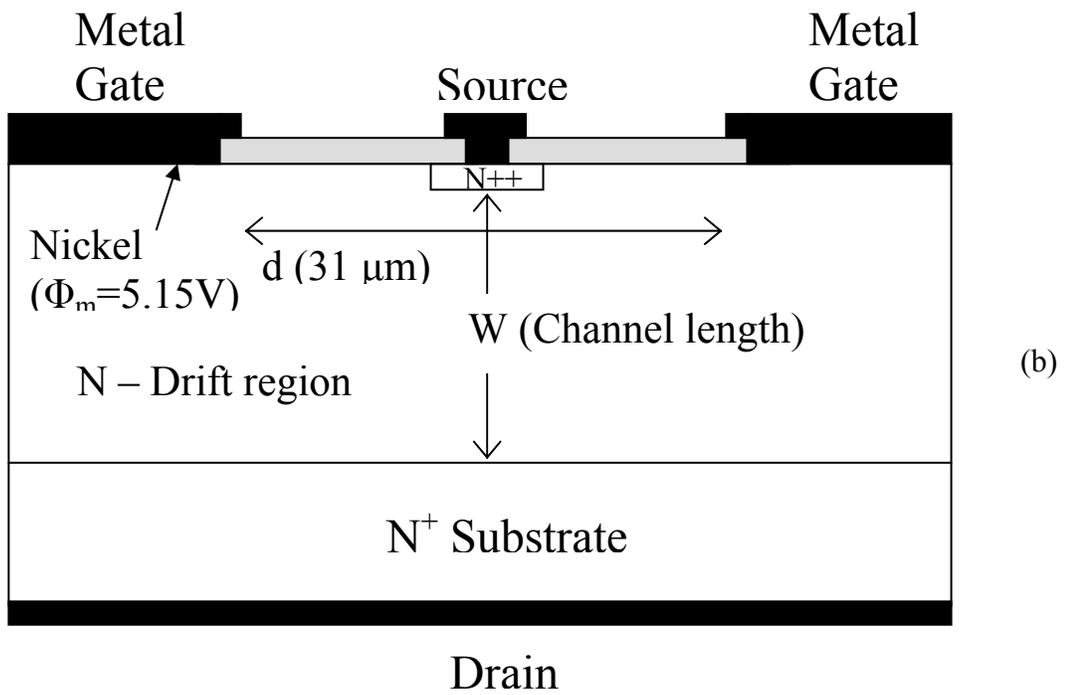

(b)

Fig.1

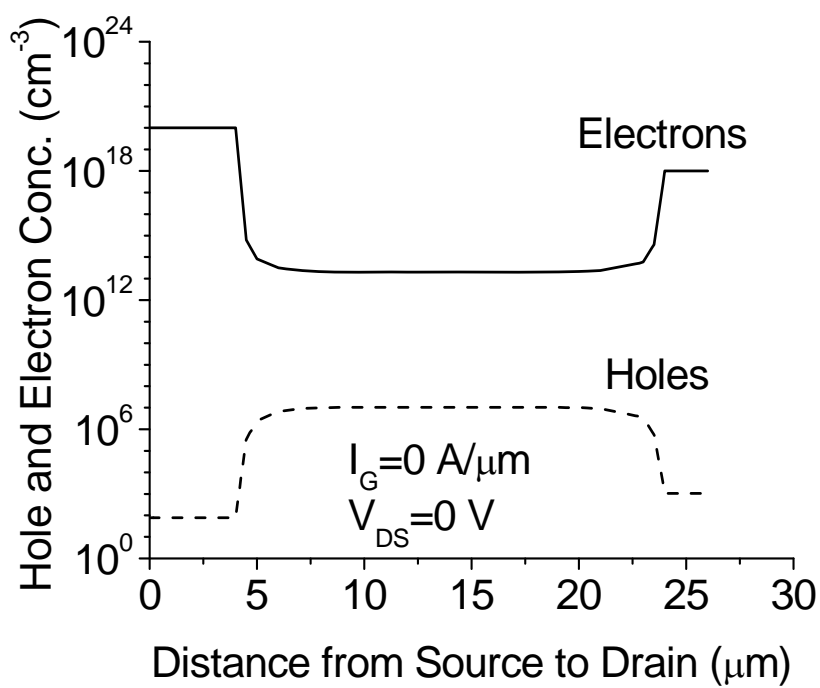

Fig. 2 (a)

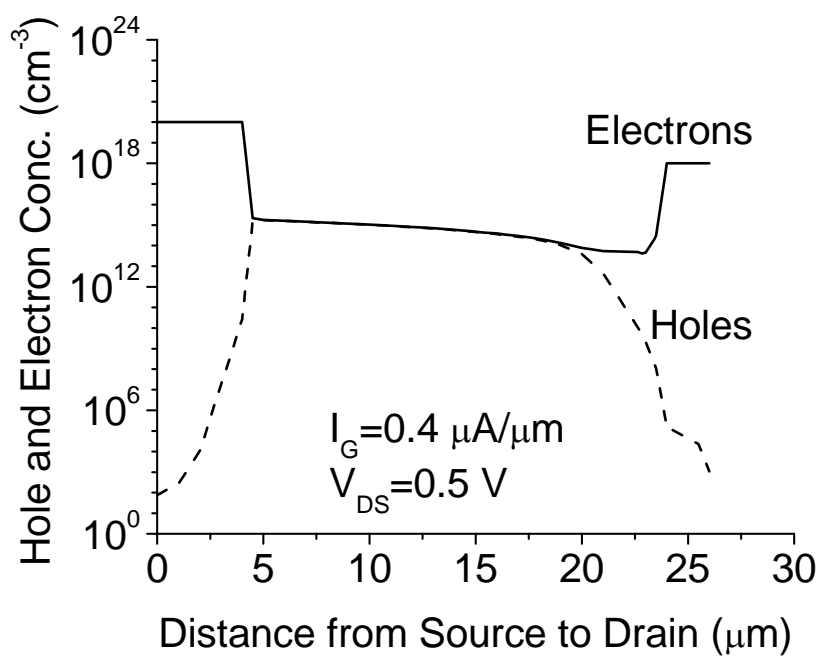

Fig. 2 (b)

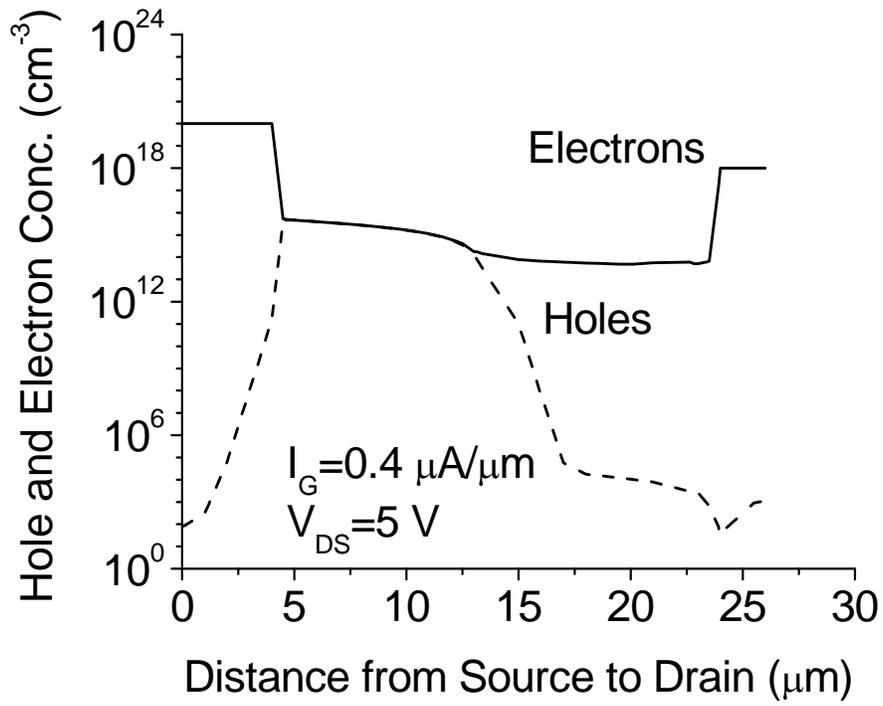

Fig. 2(c)

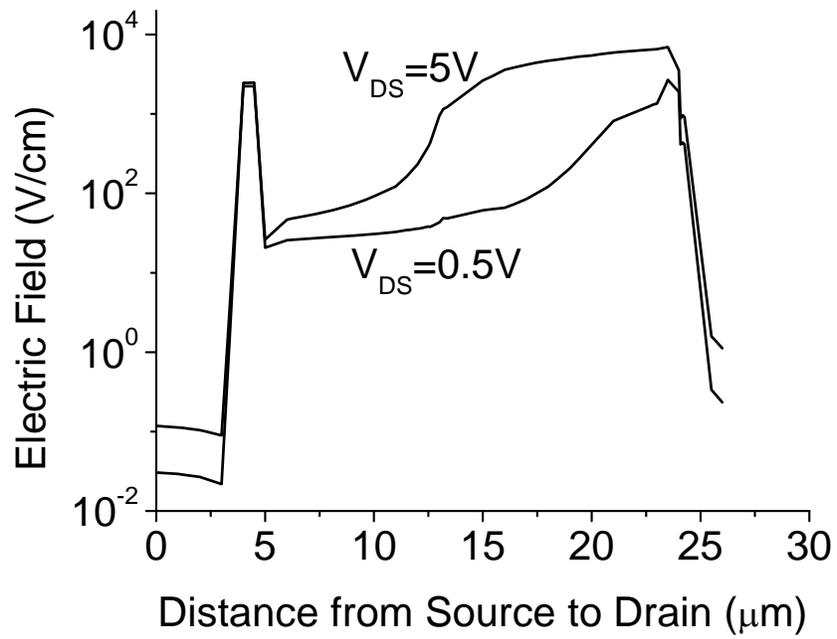

Fig. 3.

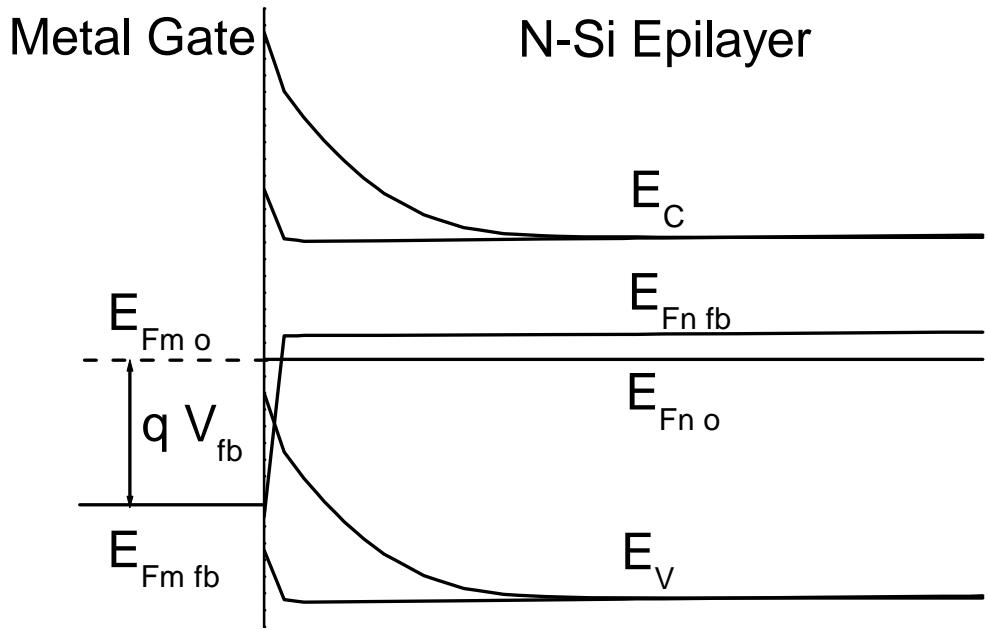

Fig. 4

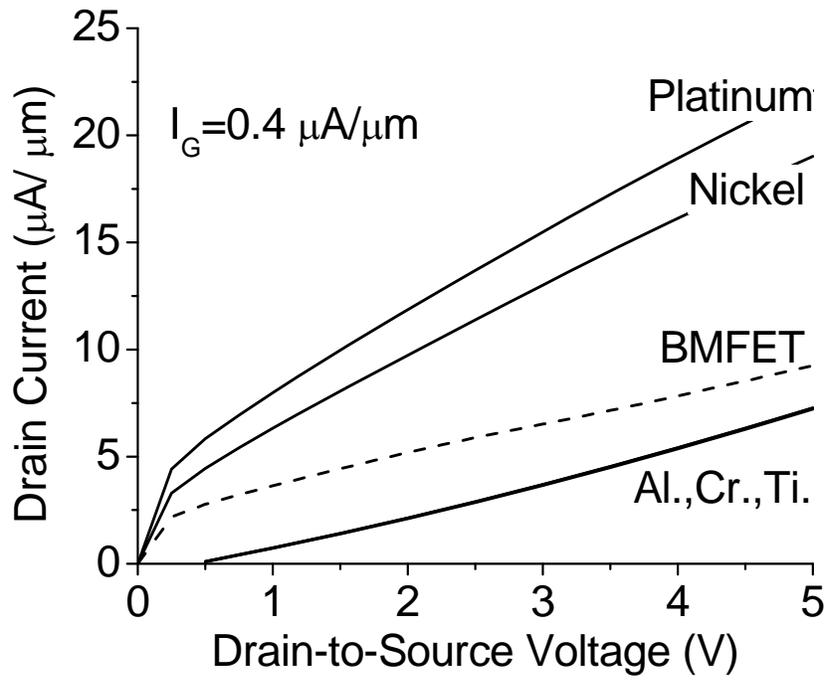

Fig. 5

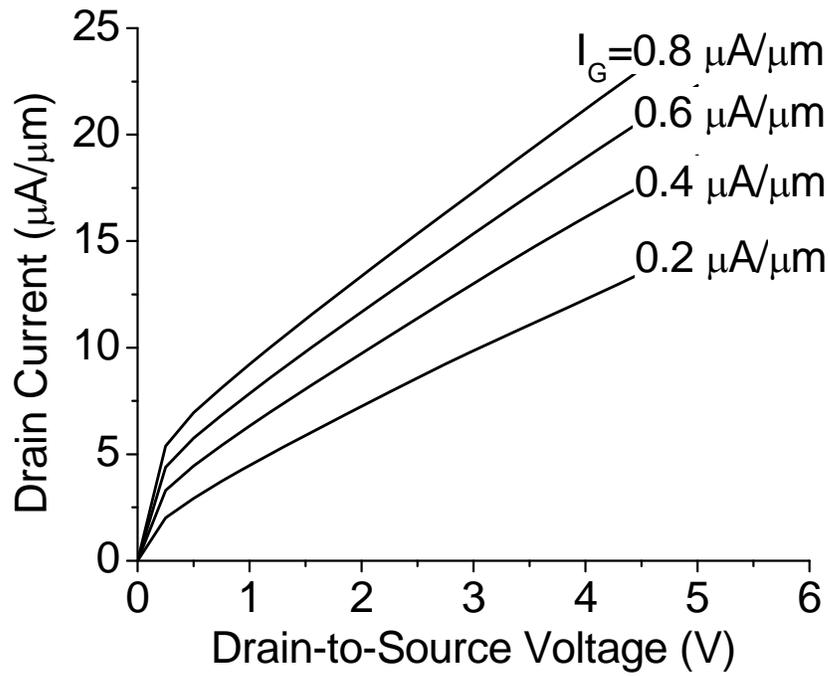

Fig. 6

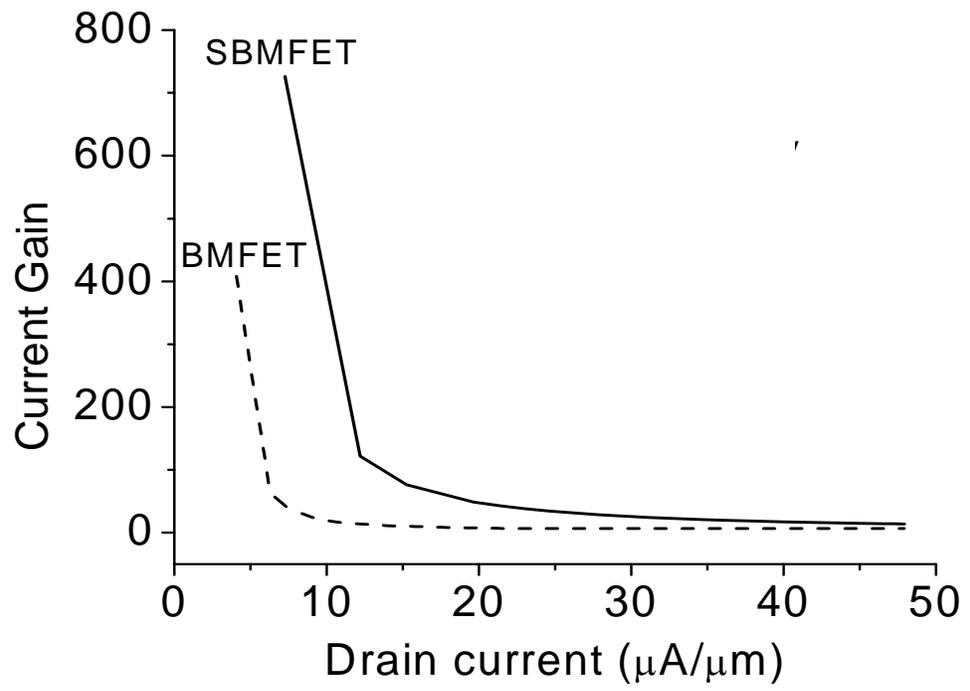

Fig. 7

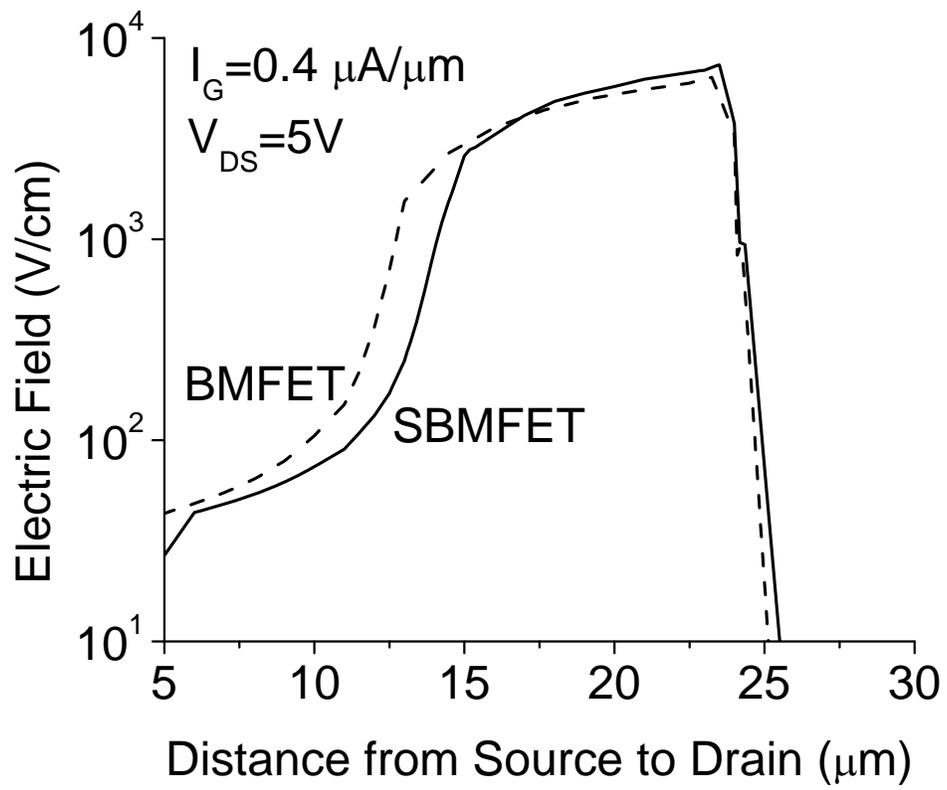

Fig. 8

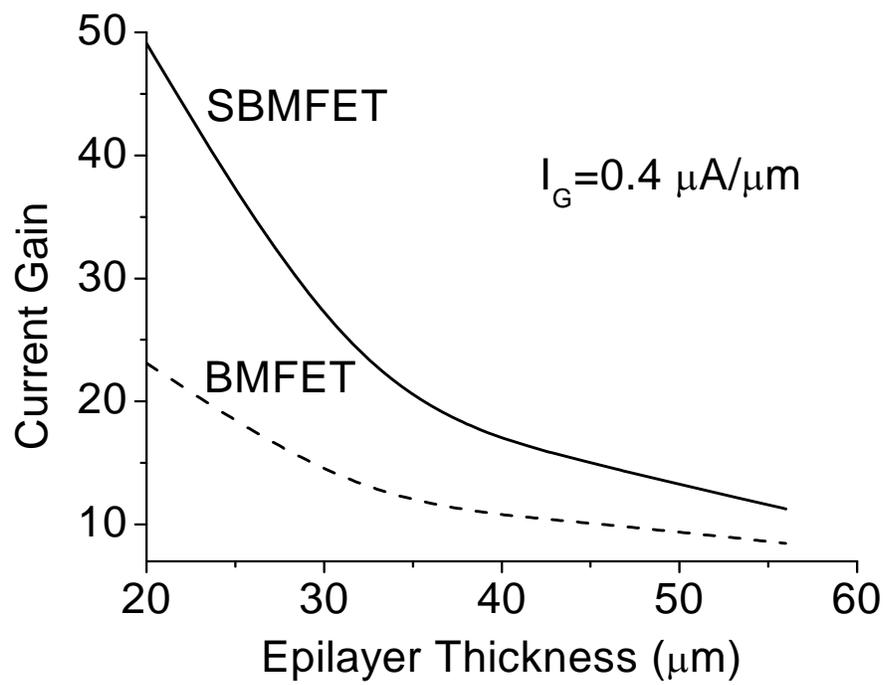

Fig. 9

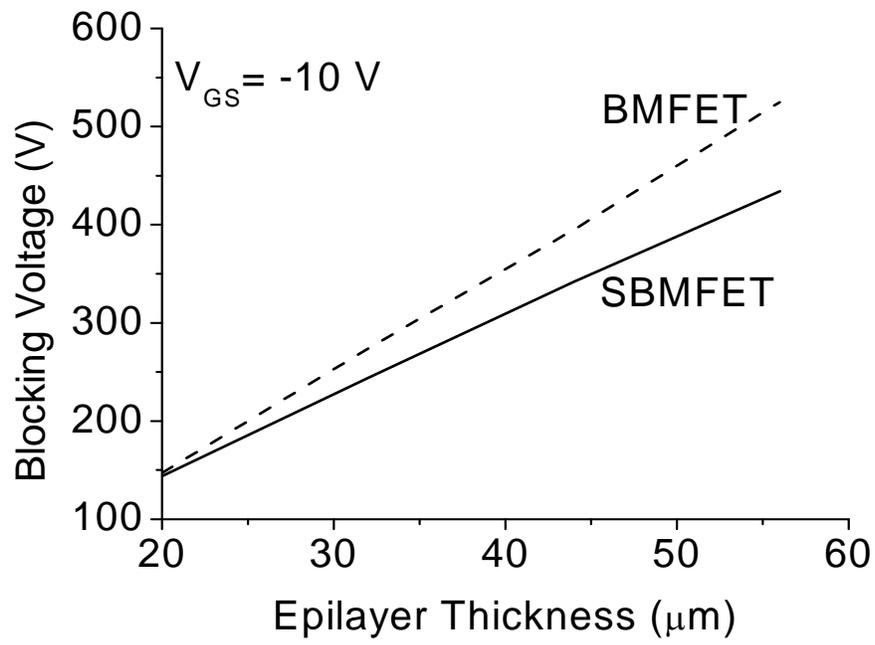

Fig. 10